\documentclass[prb,twocolumn,showpacs,preprintnumbers,amsmath,amssymb,superscriptaddress]{revtex4}

\usepackage{graphicx}
\usepackage{dcolumn}
\usepackage{bm}
\usepackage{color}
\usepackage{comment}
\usepackage{amssymb}
\usepackage{amsmath}
\usepackage{xfrac}
\usepackage{xr}
\begin{document}
\newcommand{\siox}{SiO$_2$}
\newcommand{\silicate}{Si$_2$O$_3$}
\newcommand{\sicoxint}{\mbox{SiC/SiO$_2$}}
\newcommand{\rootthree}{($\sqrt{3}$$\times$$\sqrt{3}$)R30$^{\circ}$}
\newcommand{\rootthreehl}{($\sqrt{\mathbf{3}}$$\times$$\sqrt{\mathbf{3}}$)R30$^{\circ}$}
\newcommand{\sixroot}{(6$\sqrt{3}$$\times$6$\sqrt{3}$)R30$^{\circ}$}
\newcommand{\sixroothl}{(6$\sqrt{\mathbf{3}}$$\times$6$\sqrt{\mathbf{3}}$)R30$^{\circ}$}
\newcommand{\three}{\mbox{(3${\times}$3)}}
\newcommand{\four}{\mbox{(4${\times}$4)}}
\newcommand{\five}{\mbox{(5${\times}$5)}}
\newcommand{\six}{\mbox{(6${\times}$6)}}
\newcommand{\two}{\mbox{(2${\times}$2)}}
\newcommand{\twobyc}{\mbox{(2$\times$2)$_\mathrm{C}$}}
\newcommand{\twobysi}{\mbox{(2$\times$2)$_\mathrm{Si}$}}
\newcommand{\seven}{($\sqrt{7}$$\times$$\sqrt{7}$)R19.1$^{\circ}$}
\newcommand{\fourseven}{(4$\sqrt{7}$$\times$4$\sqrt{7}$)R19.1$^{\circ}$}
\newcommand{\one}{\mbox{(1${\times}$1)}}
\newcommand{\sicbar}{SiC(000$\bar{1}$)}
\newcommand{\sicbarhl}{SiC(000$\bar{\mathbf{1}}$)}
\newcommand{\bardir}{(000$\bar{1}$)}
\newcommand{\grad}{\mbox{$^{\circ}$}}
\newcommand{\cgrad}{\,$^{\circ}$C}
\newcommand{\projecta}{(11$\bar{2}$0)}
\newcommand{\projectb}{(10$\bar{1}$0)}
\newcommand{\projectc}{(01$\bar{1}$0)}
\newcommand{\third}{$\frac{1}{3}$}
\newcommand{\thirdspot}{($\frac{1}{3}$,$\frac{1}{3}$)}
\newcommand{\oversix}{$\frac{1}{6}$}
\newcommand{\gkdir}{$\overline{\Gamma \mbox{K}}$-direction}
\newcommand{\pb}{$\pi$-band}
\newcommand{\pbs}{$\pi$-bands}
\newcommand{\kv}{\mathbf{k}}
\newcommand{\kpoint}{$\overline{\textrm{K}}$ point}
\newcommand{\kpp}{$\overline{\textrm{K'}}$}
\newcommand{\kp}{$\overline{\textrm{K}}$}
\newcommand{\gpoint}{$\overline{\textrm{\Gamma}}$ point}
\newcommand{\gk}{$\overline{\Gamma \mbox{K}}$}
\newcommand{\kpar}{$\mathbf{k}_{\parallel}$}
\newcommand{\efermi}{E$_\mathrm{F}$}
\newcommand{\edirac}{E$_\mathrm{D}$}
\newcommand{\angs}{$\mathrm{\AA}$}
\newcommand{\kk}  {$\mathbf{k}_{\overline{\textrm{K}}}$}
\newcommand{\kvec}{\underline{k}}
\newcommand{\sbs}[2]{\rlap{\textsuperscript{{#1}}}\textsubscript{#2}}
\newcommand{\round}[1]{\left({#1}\right)}
\newcommand{\moire}{moir{\'e}}
\newcommand{\gp}{$\overline{\Gamma}$}
\newcommand{\mmp}{$\overline{\textrm{M}}$}
\newcommand{\msic}{$\overline{\textrm{M}}_{\textup{SiC}}$}
\newcommand{\ksic}{$\overline{\textrm{K}}_{\textup{SiC}}$}
\newcommand{\mau}{$\overline{\textrm{M}}_{\textup{Au}}$}
\newcommand{\kau}{$\overline{\textrm{K}}_{\textup{Au}}$}
\newcommand{\mgr}{$\overline{\textrm{M}}_{\textup{Gr}}$}
\newcommand{\kgr}{$\overline{\textrm{K}}_{\textup{Gr}}$}

\newcommand{\red}{\textcolor[rgb]{1,0,0}}

\newcommand{\todo}[1]{\textsl{\textcolor{red}{#1}}}


\title{Semiconductor to metal transition in two-dimensional gold and its van der Waals heterostack with graphene}

\author{S. Forti}
\email{stiven.forti@iit.it}
\affiliation{Center for Nanotechnology Innovation @ NEST, Istituto Italiano di Tecnologia, Piazza San Silvestro 12, 56127 Pisa, Italy}
\affiliation{Max-Planck-Institut f\"{u}r Festk\"{o}rperforschung, Heisenbergstr. 1, D-70569 Stuttgart}
\author{S. Link}
\affiliation{Max-Planck-Institut f\"{u}r
Festk\"{o}rperforschung, Heisenbergstr. 1, D-70569 Stuttgart}
\author{A. St\"ohr}
\affiliation{Max-Planck-Institut f\"{u}r
Festk\"{o}rperforschung, Heisenbergstr. 1, D-70569 Stuttgart}\
\author{Y. R. Niu}
\affiliation{MAXIV laboratory, Lund University, P.O. Box 118, Lund, S-22100, Sweden}
\author{A. A. Zakharov}
\affiliation{MAXIV laboratory, Lund University, P.O. Box 118, Lund, S-22100, Sweden}
\author{C. Coletti}
\email{camilla.coletti@iit.it}
\affiliation{Center for Nanotechnology Innovation @ NEST, Istituto Italiano di Tecnologia, Piazza San Silvestro 12, 56127 Pisa, Italy}
\affiliation{Graphene Labs, Istituto Italiano di Tecnologia, via Morego 30, 16163 Genova, Italy}
\author{U. Starke}
\affiliation{Max-Planck-Institut f\"{u}r Festk\"{o}rperforschung,
Heisenbergstr. 1, D-70569 Stuttgart}

\begin{abstract}
{\bf
\noindent 
The synthesis of transition metals in a two-dimensional (2D) fashion has attracted growing attention for both fundamental and application-oriented investigations, such as 2D magnetism, nanoplasmonics and non-linear optics. However, the large-area synthesis of this class of materials in a single-layer form poses non-trivial difficulties. Here we present the synthesis of a large-area 2D gold layer, stabilized in between silicon carbide and monolayer (ML) graphene. We show that the 2D-Au ML is a semiconductor with the valence band maximum 50 meV below the Fermi level. The graphene and gold layers are largely non-interacting, thereby defining a novel class of van der Waals heterostructure. 
The 2D-Au bands, exhibit a 225 meV spin-orbit splitting along the {\gk} direction, making it appealing for spin-related applications. By tuning the amount of gold at the SiC/graphene interface, we induce a semiconductor to metal transition in the 2D-Au, which was never observed before and hosts great interest for fundamental physics.
}
\end{abstract}

\maketitle

\date{\today}


Lowering the dimensionality of a given compound has become a big hype in the field of material research, as this opens the possibility to explore and exploit properties which are inaccessible in a bulk crystal. While several compounds can be relatively easily rendered two-dimensional (2D) with physical approaches\cite{Berger2004,NovoselovScience2004,NagashimaPRL1995,RuitaoACR2015,BoubekeurAPL2010,DavilaNJP2014,BampoulisJPCM2014,ZhuNatMat2015,JiNatComm2016,
ReisScience2017,LiNanoLett2013,DuanNatComm2014,ZhaoScience2014},
many others are difficult to isolate. In particular, there are essentially no single-element materials other than graphite that can be easily exfoliated \cite{MounetNatNanotech2018}. 
2D metals see a vast field for applications, ranging from catalysis to sensing\cite{ChenChemRev2018,SinibaldiSaA2012}, passing through enhanced magnetism\cite{Freeman1987,Yafet1986,Siegmann1992,Gibertini2019}. This enticing applicative potential has prompted the development of various approaches for the synthesis of low-dimensional metals\cite{ChenChemRev2018}. Several chemistry-based strategies have been implemented, however mostly yielding nanostructures of different shapes and geometries\cite{ChenChemRev2018}. Physical methods have been shown to effectively yield monolayers of transition metals, typically onto other metal surfaces\cite{JacobsenPRL1995,ShikinPRL2008,ShikinNJP2013}. However, keeping the surface energy low enough to avoid the growth of multilayer islands is a non-trivial problem to address and often a capping medium solution must be adopted\cite{LeeNatRevMat2016}.\\ In all these conventional surface science studies, the main focus has been put in investigating the effects of the low dimensional metal on the substrate electronic properties, with the metal adlayer typically causing large spin-orbit splitting in the substrate band structure\cite{ShikinNJP2013}. However, the strong interaction between the metallic substrate and the 2D layer has hindered the possibility to access the intrinsic properties of the latter.\\ 

In this article, we shed a light onto the electronic properties of two-dimensional gold (2D-Au), by synthesizing it via the intercalation of Au atoms at the heterointerface between the Si-terminated SiC(0001) surface and its C-rich {\sixroot} reconstruction. Au atoms are arranged in a highly crystalline fashion and develop their own band dispersion, which is intrinsically two-dimensional. When a single layer of Au atoms is intercalated, a semiconductor (SC) crystal with the valence band maximum in {\kau} close to the Fermi level and a saddle point in {\mmp} at about 400 meV is formed. These 2D states also exhibit a significant spin-orbit splitting, amounting to 225 meV, along the {\gk} dispersion direction. We show that the 2D semiconducting Au undergoes a transition to metal (M) when the number of intercalated Au layers goes from one to two. This is the first experimental report of a semiconductor-to-metal transition in a 2D transition-metal thus hosting great interest at a fundamental level.\\ 

Transition metal atoms have already been intercalated underneath the buffer layer graphene~\cite{EmtsevPRB2008,RiedlJPD2010,FortiJPD2013} or zerolayer graphene (ZLG) on SiC(0001) as well as under epitaxial graphene on Ni(111)\cite{Varykhalov2008}. The effects reported were doping of the graphene {$\pi$}-bands\cite{GierzNL2008,GierzPRB2010} or mini-gaps opening in the graphene Dirac cone due to {\moire} superperiodicity\cite{Forti2DMat2016}. It was recently shown that the interaction between the 5d orbitals of gold and the valence band of graphene can open a spin-orbit gap deep in the valence band of gold-intercalated graphene on SiC(0001)\cite{MarchenkoAPL2016}. However, no evidence of two-dimensional dispersing bands stemming from the metal layer was ever reported, which would be indicative of a well-defined order of the film.\\

\begin{figure*}[htb]
\centering
\includegraphics[width=0.95\textwidth]{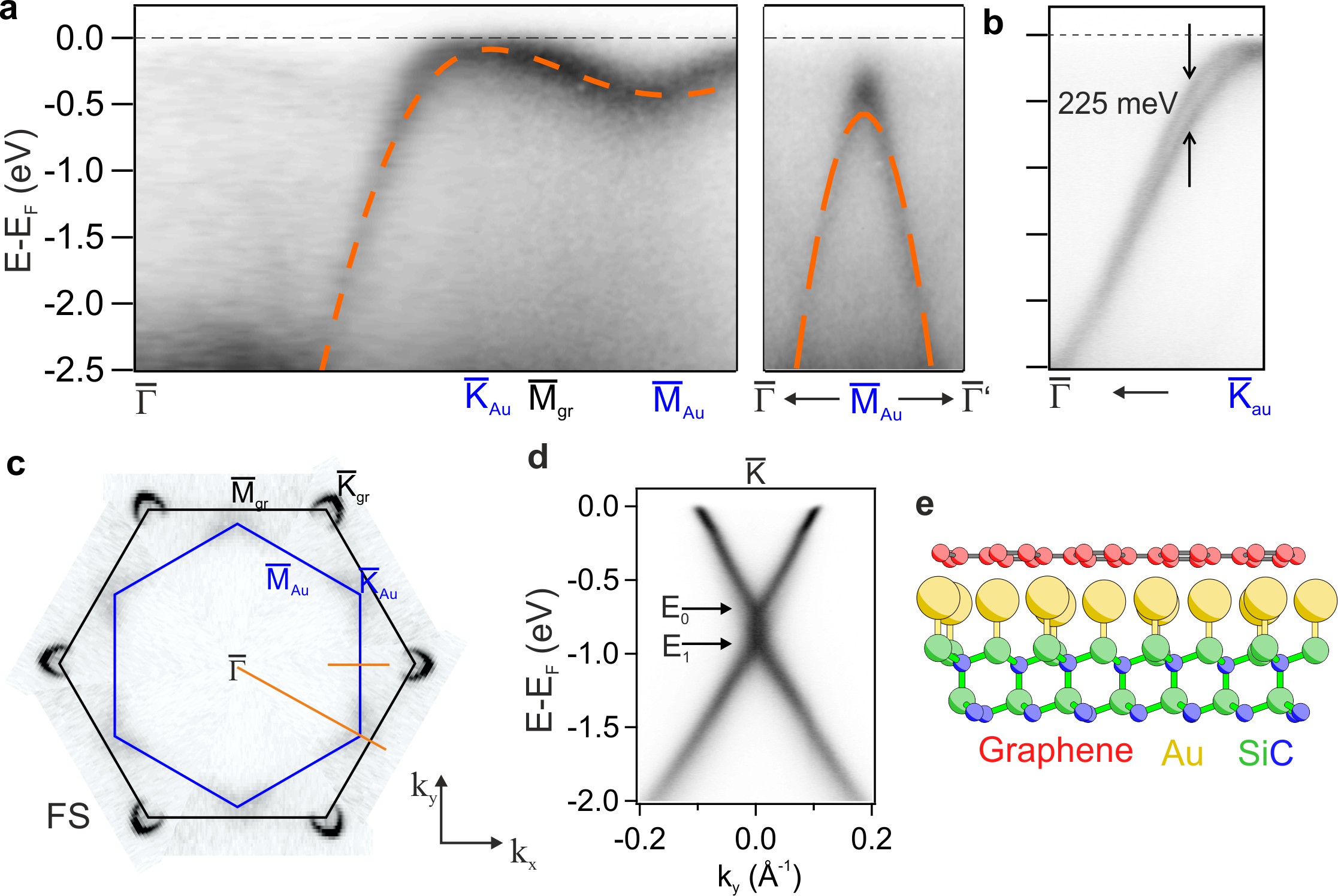}
\caption{{\bf Electronic properties of 2D Gold.} (a) ARPES cut along the diagonal orange line in panel (d), showing only the Au bands dispersion. In the right panel the spectrum cuts along the short orange line in panel (d). The orange dashed lines superimposed to the data are the TB bands as described in the text. (b) Au bands dispersion along the $k_y$ direction highlighting their spin-orbit splitting $\Delta SO$. The photon energy for panels (a) and (b) is 95 eV. (c) Fermi surface of the system measured by ARPES. The figure is the result of the sixfold symmetrisation of a measurement. (d) Dispersion measured along $k_y$ at the {\kgr} point with 40 eV photons, showing graphene's Dirac cone. (e) Ball-and-stick model of the gold atoms intercalated between graphene and SiC(0001).}
\label{Fig1}
\end{figure*}

Figure\ref{Fig1}(a) and (b) displays the band structure of the SC 2D-Au, measured by angle-resolved photoemission spectroscopy (ARPES). Two spectral cuts are shown in Fig.~\ref{Fig1}(a), which give the dispersion along the high symmetry directions {\gp\kau}{\mau} and {\gp}{\mau}{\gp}'. From {\gp} towards {\kau}, the dispersion has a band slope of 6 eV{\AA} at a binding energy of about -1 eV, i.e. comparable with graphene's band velocity in the vicinity of the Fermi level. From {\kau} towards {\mau}, the band is instead rather flat. At {\kau} we identify the maximum located at -50 meV. By performing a statistical analysis on the position of the valence band maximum (VBM) of 2D-Au for several samples, we find that the overall mean value falls just short of -70 meV (cf. sec. $I$ of the supplementary information (SI)). The saddle point in {\mau} is found at -400 meV, meaning that the van Hove singularity (vHS) in the density of states is at an energy of interest for electronic measurements. No distinct crossing of the Fermi level is observed, in line with the missing Fermi surface contour in Fig.~\ref{Fig1}(c). We observe a faint triangularly shaped spectral weight at the Fermi level, due to the intrinsic width of the band. By connecting the vertexes of the aforementioned intensities, we define the Brillouin zone (BZ) of the SC 2D-Au, drawn as a blue hexagon in Fig.~\ref{Fig1}(c).
Photon energy dependent ARPES measurements were performed to probe a possible crystallographic periodicity along the $\hat{z}$ axis (see sec.~$II$ of the SI), and confirmed the two-dimensional nature of the gold layer. Furthermore, a high-resolution close-up of the bands along the {\gk} direction shows that the states exhibit a spin-splitting which reaches the maximum value of 225 meV at -0.5 eV. \\
It should be mentioned that since the states observed in this energy range and in this $k$-space portion have no correspondence in any other graphene/SiC(0001) based system~\citep{BostwickNP2007,EmtsevPRB2008,Forti2DMat2016} they can solely be attributed to the interfacial Au layer.

The gold atoms at the interface affect the electronic properties of graphene \citep{GierzPRB2010}. A single gold layer act as an electron donor to graphene, shifting the {\pbs} down until the Dirac point $E_0$ is observed about 700 meV under the Fermi level, as visible from Fig.\ref{Fig1}(d), where we show an ARPES cut through graphene's {\kgr} point, taken perpendicular to the {\gp\kgr} direction. In addition, the overall dielectric constant of the system is such that many-particle effect emerge and can be detected. In particular, we extract an effective dielectric constant of $\epsilon_{eff} = 7\pm1$, which allow us to detect the plasmaron band\cite{BostwickScience2010}. The crossing energy of the plasmaron band is indicated as $E_1$ and further details can be found in the sec. $III$ of the SI.     

The size and shape of the 2D-Au BZ suggest that the Au atoms are arranged on a triangular lattice with the lattice parameter matching that of SiC(0001), as it is depicted in the ball-and-stick model of Fig.~\ref{Fig1}(e). To further corroborate this hypothesis, we have observed with ARPES graphene replica bands shifted by a SiC(0001) reciprocal lattice vector (cf. sec. $IV$ in SI). \\
We compared the observed bands to a tight binding (TB) model calculated on this 2D triangular lattice in the next-nearest neighbor (NNN) approximation:
\begin{eqnarray}
  g(\vec{k}) &=& -\varepsilon_0 - \gamma_{01}u_{1}(\vec{k}) - \gamma_{02}u_{2}(\vec{k}) \\
  u_{1}(\vec{k}) &=& 2\cos(k_x a) + 4\cos(k_x a/2)\cos(k_y \sqrt{3}a/2) \nonumber\\
  u_{2}(\vec{k}) &=& 2\cos(k_y \sqrt{3}a) + 4\cos(k_y \sqrt{3}a/2)\cos(k_x 3a/2) \nonumber
  \label{Eq1}
\end{eqnarray}
where $a$ is the lattice parameter, $\gamma_{01}$ and $\gamma_{02}$ are the NN and NNN hopping parameters, the first being of the order of 1 eV and $\gamma_{02}\simeq\gamma_{01}/10$. For the plot of Fig.~\ref{Fig1}(a) and (b) we set $\gamma_{01}=1.3$ eV and $a=3.08$ {\AA}, that is the SiC(0001) lattice parameter.

Apparently, even this simple model describes the measured band dispersion quite well. In particular, the extremes are placed at the same $k$, confirming the right choice of $a$. The validity of the model also confirms the 2D nature of the states. However, some small deviations are present, especially at {\mau}. 

Density functional theory (DFT) calculations have been published for this system by Chuang \textit{et al.}~\cite{ChuangNanotech2011}. These simulations show the formation of Au-interface related bands, similar to the ones observed in our experiment. There, it was presumed, that the structure of the interfacial Au is arranged such that every topmost Si atom of the SiC is bound to one Au atom, i.e. a (1$\times$1) with respect to the SiC(0001). Similar to our experiment, their bands have the maximum in close proximity to the Fermi level, which indicates a semiconducting behavior.\\
Moreover, in Ref.\cite{HsuAPL2012} the authors find a strong spin-splitting stemming from spin-orbit coupling. The spin texture there resembles a Rashba-type with a Rashba parameter of 1.638 eV/{\AA}. In fact, we do find a strong correspondence to this in the experimental band structure (see Fig.~\ref{Fig1}(b)). Extracting the Rashba parameter from the experiments, we find a value of 2.17 eV/{\AA}, which is in rather good agreement with theory.\\

\begin{figure}[t!]
\centering
\includegraphics[width=0.5\textwidth]{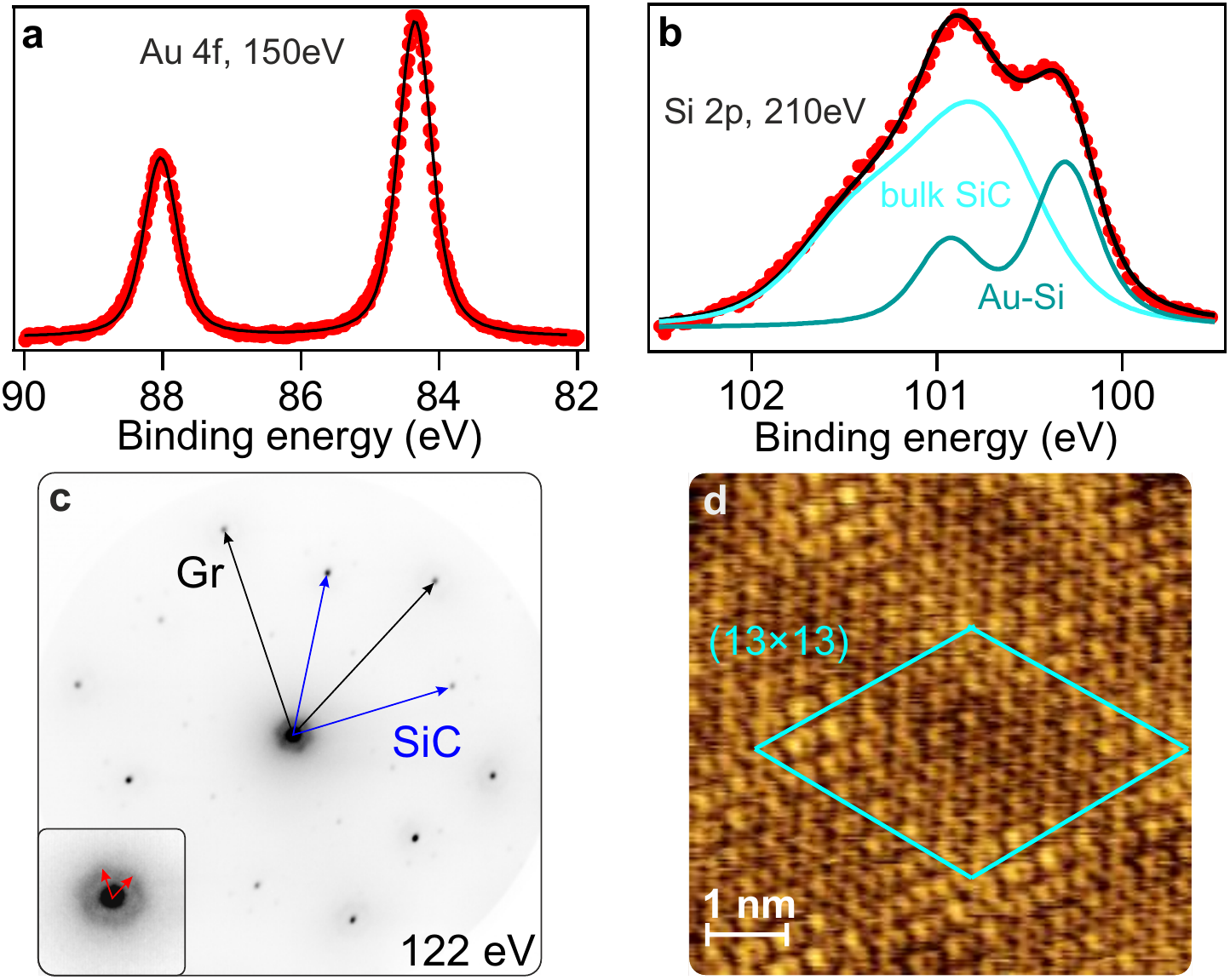}
\caption{{\bf Structural properties of the interface.} Core level spectra acquired on a Au-intercalated monolayer of epitaxial graphene on SiC(0001) of (a) Au 4f fitted with a single Voigt doublet, (b) Si 2p fitted with two Voigt doublets. The photon energies are indicated in the panels. (c) $\mu$LEED pattern recorded at 122 eV on Au-intercalated graphene. The inset highlights the (13$\times$13) pattern observed around the (00) spot. (d) STM topography image over a (5.2$\times$5.2) nm$^2$ region. The (13$\times$13) {\moire} unit cell is indicated as a cyan diamond. Image recorded at U$_{sample}$=100 mV and I$_{tunnel}$=85 pA.}
\label{Fig2}
\end{figure}

In order to gather further information about the actual structure of the interfacial gold layer, we carried out high-resolution X-ray photo-emission spectroscopy (HRXPS), low-energy electron diffraction (LEED) and scanning tunneling microscopy (STM) measurements. Representative spectra of the Au~4f and Si~2p core levels are shown in Fig.~\ref{Fig2}(a) and (b), respectively.
The details about the fitting parameters are reported in section~$V$ of the SI. The Au 4f doublet indicates the presence of atoms in a single chemical environment. The binding energy that we measure is $\sim$350 meV higher than for metallic gold and this excludes the presence of a substantial amount of metallic gold either on the surface or underneath graphene. In contrast, it indicates that Au is bound to Si at the interface, as confirmed by the Si 2p spectrum, which shows a component of bulk SiC and a surface component of Au-Si. 
Symmetry considerations allow us to conclude that, since the bands we observe are of $s-p$ character, the Au atoms must interact with the 2$p$ Si orbitals by their 5$d$ electrons.
It should be noted that this does not represent an alloyed silicide layer, which is also reported to have a different (higher) binding energy\cite{ChariaSS2006}. What we observe is rather a significant electronic overlap between the topmost Si atoms of the substrate and the intercalated gold layer. The interaction with the substrate stabilizes the gold layer and induces the interfacial order. Notably, Au atoms deposited onto SiC(0001) are observed not to order on the (1$\times$1)\cite{ChariaSS2006,StoltzJPCM2007}, which implies the need for the graphene on top to impose the observed ordering and electronic structure.\\

\begin{figure*}
\centering
\includegraphics[width=0.99\textwidth]{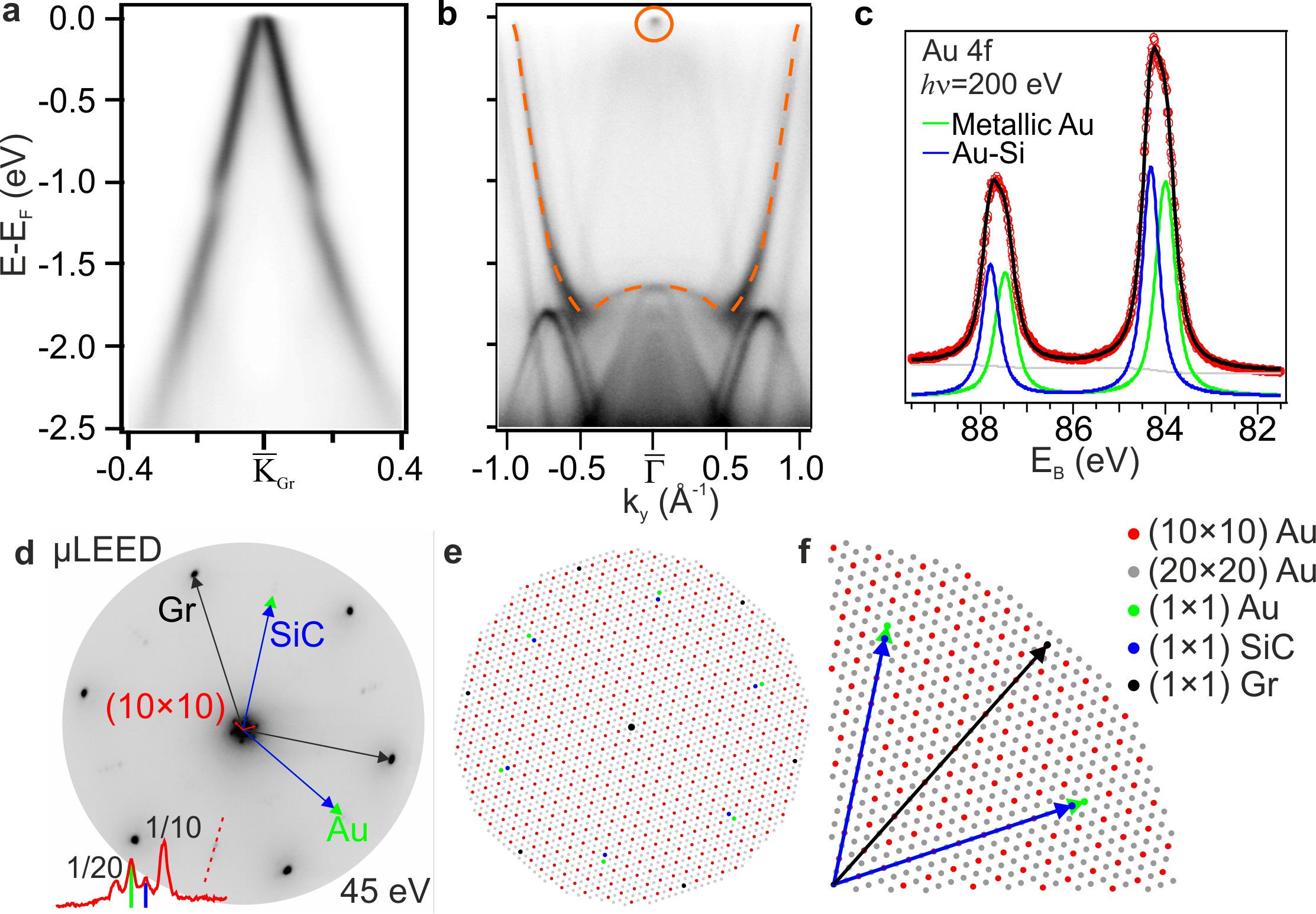}
\caption{{\bf Properties of the $p$-phase Au-intercalated graphene.} (a) ARPES spectrum of the graphene's Dirac cone measured along k$_y$, recorded with photons of 40 eV. (b) ARPES spectrum measured at 75 eV centered in {\gp}, showing the band structure of the intercalated gold along the k$_y$ direction. (c) $\mu$XPS of the system of Au 4f doublet. (d) $\mu$LEED pattern of the system. In the bottom-left corner the line profile along the red line is shown, highlighting the four diffraction spots visible and assigning them to SiC (blue), Au (green) and the 1/20 and 1/10 Au superperiodicities. (e) sketch of the $\mu$LEED with all the periodicities indicated. (f) blow-up of one quadrant of panel (e).}
\label{Fig3}
\end{figure*}
From the previously reported HRXPS data, we find that every Au atom is bound to a Si atom at the interface (cf. SI), meaning that the Au is ordered in a (1$\times$1) registry with respect to the SiC, as it was found for ARPES. The $\mu$-LEED pattern of Fig.~\ref{Fig2}(c) shows that after the intercalation of Au atoms, only the graphene and SiC (1$\times$1) are visible. In addition, a hexagonal halo oriented like graphene and 1/13 of its dimension is discernible, corroborating the proposed model. Atomically resolved STM carried out on such a sample, further confirms the atomic model. The STM image in Fig.~\ref{Fig2}(d) shows a region where the (13$\times$13) superpotential is well visible and the graphene lattice is atomically resolved (see also sec. $VI$ of the SI).


By extending the gold deposition time, it is possible to intercalate a double layer of gold (see Methods for more details). Going from a single Au layer to a Au bilayer does not only switch the doping of graphene from $n$+ to slightly $p$, but it also switches the character of the 2D-Au electronic properties. In fact, we observe a transition from SC to a M behavior.

Fig.~\ref{Fig3} summarizes the electronic, chemical and structural properties of the M 2D-Au. In the first place, the graphene {\--} shown in panel {\bf a} {\--} is marginally hole-doped with a density of states of (9$\pm$1)$\cdot$10$^{11}$cm$^{-2}$, which, given a band velocity of (1.33$\pm$0.05)$\cdot$10$^{6}$ m/s, corresponds to a shift of the Fermi level by (-150$\pm$10) meV with respect to the Dirac point. We as well observe the Rashba splitting at about -1 eV due to the interaction between the graphene and gold bands, as reported in Ref.\cite{MarchenkoAPL2016}.\\
The interfacial gold still develops its own band dispersion, with a rich structure, as exemplary shown in Fig.~\ref{Fig3}(b), where we show an ARPES spectrum recorded at 75 eV and centered in {\gp} along k$_y$ direction, with reference to Fig.~\ref{Fig1}(c). We indicate the states reaching the Fermi level with the orange dashed line and a solid circle. The spectral weight at {\gp} at the Fermi level represents the bottom of a conduction band of the M 2D Au and it is not related with the Shockley state of the Au(111) surface. It was predicted in previously published calculations \cite{ChuangNanotech2011}, even if they considered three layers at the interface instead of two. Such a band structure implies that at any given small variation of the chemical potential, or applied external voltage, there will always be a finite electronic density of states, confirming the metallicity of this phase.\\
HRXPS measurements confirm the presence of a gold double layer, as does STM (see sec. $VI$ in the SI). Fig.\ref{Fig3}(c) displays the Au 4f doublet fitted with two components. The ratio between the two is very close to one, indicating a symmetric distribution of the atoms between the first and the second layer.\\
Due to the favored interatomic forces, the intercalated gold bilayer shrinks its lattice parameter with respect to the monolayer. The gold atoms then arrange in a configuration closer to the bulk Au(111). This is illustrated by the $\mu$-LEED data shown in Fig.~\ref{Fig3}(d). A hexagonal pattern around the (00) spot is well visible. That vector is 1/10 of the Au reciprocal lattice vector. Yet, even a denser grid is visible, as the line profile (enlarged to be better seen) in the bottom left part of the panel indicates. \\
Such a grid emerges from the Au/SiC(0001) interface and it corresponds to Au atoms being arranged as (20$\times$20) over (19$\times$19) SiC unit cells. The corresponding lattice parameter for the gold M-phase is extracted from several different LEED, $\mu$LEED, ARPES and STM measurements. The estimation of the lattice parameter is (2.93$\pm$0.01) {\AA}, about 1.6\% larger than the nominal 2.883 {\AA} of the Au(111).

\begin{figure}[t]
\centering
\includegraphics[width=0.5\textwidth]{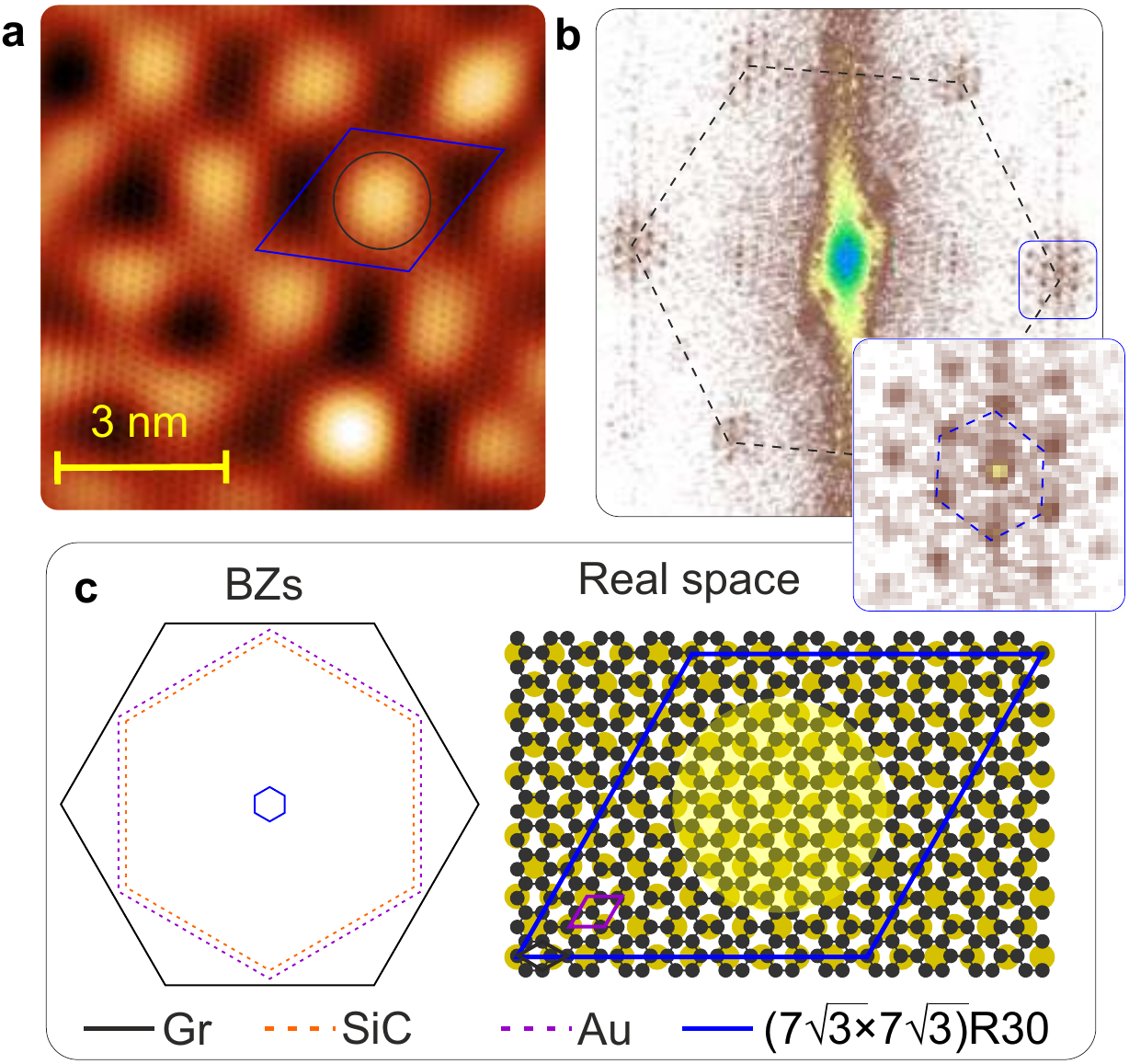}
\caption{(a) STM image 2D-FFT filtered U$_{sample}$=-105 mV I$_{tunnel}$=325 pA. (b) 2D-FFT of the real-space image in panel (a). In the inset: zoom-in with respect to the blue box in panel (b). (c) reciprocal and real-space arrangement between graphene and gold. All BZs are 90{$^\circ$} rotated in order to match the alignment used for the ARPES description.}
\label{Fig4}
\end{figure}

When such a gold bilayer interacts with graphene, it develops a superperiodicity half the size, which is observed in STM, as visible in the data reported in Fig.~\ref{Fig4}. The Gr/Au {\moire} is well identified in the topographical STM measurements of Fig.~\ref{Fig4}(a). The graphene lattice is very well resolved as well and the superlattice unit cell is drawn in the image as a blue diamond. In Fig.~\ref{Fig4}(b), the fast 2D Fourier transformation (2D-FFT) of the image (raw data in Fig. S7 of the SI) shows distinctively the graphene and superstructure periodicities and helps us conclude that (7$\sqrt{3}\times$7$\sqrt{3}$)R30 graphene unit cells are arranged on (10$\times$10) of gold, which is the strongest superperiodicity observed also in LEED (cf. Fig. S8 in the SI). The proposed model for the atomic arrangement of graphene and M 2D Au is shown in Fig.~\ref{Fig4}(c). The region within the supercell where the C-Au distance is larger is circled in panel (a) and indicated as a yellow disc in panel (c). The difference in C-Au adsorption distance emerges from the local variation of the atomic registry between the lattices. 
 
In summary, we presented the first experimental realization of a vdWH between a 2D semimetal and a 2D semiconductor, where the semiconductor is realized by synthesizing a single layer of gold at the heterointerface between graphene and 6H-SiC(0001). We have determined that the gold atoms arrange themselves in a highly ordered triangular lattice mimicking the SiC(1$\times$1) and develop their own band structure. Such an electronic band dispersion has a truly two-dimensional character and it is notably well reproduced by a simple TB model.
The transition of the 2D Au from SC to M is layer-dependent and it can be controlled by tuning the amount of intercalated gold. Single and double 2D Au layers have different electronic and structural properties and the graphene above is affected as well. The system therefore results in a double lateral junction: the graphene exhibit a $p$/$n^{+}$ junction, in correspondence to the M/SC transition region of the 2D Au, which represents a lateral Schottky contact. In addition, the system has a vHS at about 400 meV from the Fermi edge, making this system a very appealing platform for the observation of collective phenomena at low temperatures. The whole structure is monolithic and and does not require any transfer.\\
Besides the deep fundamental interest for a new class of materials, we envision a vast portfolio of possible applications for this system, ranging from detection of THz radiation, to more sophisticated optoelectronic devices involving the spin degree of freedom. It will moreover serve as a platform for exploring non-linear optics effects in two-dimensional gold. A field so far essentially unexplored.



\section{Acknowledgements}

This work was supported by the Deutsche Forschungsgemeinschaft in the framework of the Priority Program 1459 Graphene (Sta315/8-2). This research
was partially funded by the European Community's Seventh Framework Programme:
Research Infrastructures (FP7/2007-2013) under grant agreement no. 226716.
Support by the staff at MAX-Lab (Lund, Sweden) is gratefully acknowledged. The research leading to these results has received funding from the EuropeanUnion's Horizon 2020 Research and Innovation Program under grant agreement No. 785219{\-}GrapheneCore2. A.A.Z. would like to acknowledge support from Siftelsen f{\"o}r Strategisk Forskning (project RMA15-0024). A special thank goes to Thiagarajan Balasubramanian for his invaluable help for the acquisition of the ARPES spectra at the I4 beamline of the MAXlab. \\
We thank HZB (Berlin, Germany) as well as MAX-lab (Lund, Sweden) for the allocation of synchrotron radiation beamtime. Support by the staff at Bessy (HZB) and MAX-Lab is gratefully acknowledged.

\section{Author contributions}
S.F. designed the experiments, grew the samples, performed ARPES, XPS and STM measurements, carried out the analysis of the data and wrote the manuscript. S.L. and A.S. helped with the measurements, the data analysis and the writing of the manuscript. Y.N. and A.Z. carried out LEEM/PEEM measurements. C.C. participated to and supervised the STM measurements. U.S. participated to the synchrotron measurements and co-designed the experiments. C.C and U.S. supervised the project. All authors edited and commented the manuscript.

\section{Methods}

\subsection{Sample preparation}
Nominally on-axis oriented single crystalline single-side polished 6H-SiC $n$-type substrates were used in the present study and were purchased from SiCrystal GmbH. 
Polishing damages were removed by etching the substrate at 1500 {\cgrad} in a purified H$_2$ atmosphere.
The SiC(0001) surfaces were graphitized by annealing the samples in argon atmosphere following the procedure introduced by Emtsev \textit{et al.} in Ref.~\cite{Emtsev2009}. The buffer layer or zerolayer graphene (ZLG) develops upon annealing in Ar atmosphere at about 1400~{\cgrad} for 10 minutes \cite{vanBommelSS1975,ForbeauxPRB1998,MattauschPRL2007,VarchonPRL2007,RiedlPRB2007,EmtsevPRB2008}. The formation of the ZLG results in a carbon-rich reconstruction with periodicity {\sixroot} with respect to the SiC(0001) unit cell\cite{RiedlJPD2010,MRSBull2012,FortiJPD2013}.
The samples were prepared by evaporating atomic Au from a Knudsen cell onto the ZLG surface, keeping the sample at temperatures between 600 and 700 {\cgrad} and by subsequent annealing of the sample at temperatures between 800 and 850 {\cgrad}. Keeping the sample at that temperature should favor the intercalation process over the diffusion of Au atoms over the surface and the consequent formation of clusters. Au is known to have a very high mobility on graphene\cite{LiuCrystals2013} so that deposition at room temperature followed by annealing leads to cluster formation\cite{GierzPRB2010}. Preparing the samples in the way described here helps in keeping the surface more clean. The single layer gold is prepared in the aforementioned way by depositing a nominal amount of about 20 gold monolayers at a rate of 5 {\AA} per minute. In this way the deposition takes about 10 minutes. For the gold bilayer a deposition time ranging from 20 to 25 minutes is used, employing the same rate. 
We point out that the intercalated samples are very stable in air ($\tau\gg$months) and that they do not require any particular protection from the atmospheric environment. In UHV conditions, they can be heated up to 850 {\cgrad} before the intercalated gold starts to be deintercalated and desorbed from the surface. 

\subsection{Measurements}
The band structure of the system was mapped by means of angle-resolved photoemission spectroscopy (ARPES) at the end-station 1$^2$ of the BESSY II (Berlin, Germany) synchrotron facility at a temperature of 90 K. The hemispherical electron analyzer used for the data harvesting was a Scienta R8000 with a diameter of 200 mm and an ultimate energy resolution of 1 meV. In our experiments the instrumental resolution was set to about 2.5 meV, hence smaller than the thermal broadening of the bands, i.e. about 7.5 meV. Additional ARPES measurements have been carried out at the I4 end-station of the MAXlab (Lund, Sweden) synchrotron facility with a SPECS Phoibos 100 analyzer. The aforementioned lines have also been used to acquire high-resolution X-ray photoelectron spectroscopy (HRXPS) data.
Preliminary studies to clarify the preparation procedures and to assess the sample's quality were carried out in the home-laboratory at the MPI Stuttgart using a Specs Phoibos 150 analyzer and monochromatized He II radiation for the photoexcitation.
Low-energy electron microscopy (LEEM) measurements, as well as micro-ARPES ($\mu$ARPES), micro-LEED ($\mu$LEED) and micro-XPS, were acquired at beamline I311 beamline of the Maxlab using an ElmitecIII microscope. More details about the performances of the microscope can be found elsewhere\cite{FortiPRB2011,FortiJPD2013}.
The scanning tunneling microscopy (STM) measurements were performed with an Omicron LT-STM microscope at the CNI@NEST in Pisa at a temperature of 78 K.

\newpage

\bibliographystyle{./SFbibstyle_NPG}
\bibliography{./Biblio}

\end{document}


\newcommand{\sixroot}{(6$\sqrt{3}\times 6\sqrt{3}$)R30$^{\circ}$}
\newcommand{\gkdir}{$\overline{\Gamma \mbox{K}}$-direction}
\newcommand{\pb}{$\pi$}
\newcommand{\pbst}{$\pi^{*}$}
\newcommand{\pbs}{$\pi$-bands}
\newcommand{\kv}{\mathbf{k}}
\newcommand{\kpoint}{$\overline{\textrm{K}}$ point}
\newcommand{\kpp}{$\overline{\textrm{K'}}$}
\newcommand{\kp}{$\overline{\textrm{K}}$}
\newcommand{\kps}{$\overline{\textrm{K}_{\textrm{s}}}$}
\newcommand{\kpps}{$\overline{\textrm{K}_{\textrm{s}}\textrm{'}}$}
\newcommand{\km}{$\overline{\textrm{KM}}$}
\newcommand{\gpoint}{$\overline{\textrm{\Gamma}}$ point}
\newcommand{\gk}{$\overline{\Gamma \mbox{K}}$}
\newcommand{\kpar}{$\mathbf{k}_{\parallel}$}
\newcommand{\efermi}{E$_\mathrm{F}$}
\newcommand{\edirac}{E$_\mathrm{D}$}
\newcommand{\angs}{$\mathrm{\AA}$}
\newcommand{\kk}  {$\mathbf{k}_{\overline{\textrm{K}}}$}
\newcommand{\sbs}[2]{\rlap{\textsuperscript{{#1}}}\textsubscript{#2}}
\newcommand{\round}[1]{\left({#1}\right)}
\newcommand{\moire}{moir{\'e}}
\newcommand{\gp}{$\overline{\Gamma}$}
\newcommand{\mpp}{$\overline{\textrm{M}}$}
\newcommand{\kvec}{$\mathbf{k}$}
\newcommand{\qvec}{$\mathbf{q}$}
\newcommand{\mvec}{$\mathbf{m}$}
\newcommand{\gvec}{$\mathbf{g}$}
\newcommand{\svec}{$\mathbf{s}$}
\newcommand{\emini}{E$_\mathrm{m}$}
\newcommand{\red}{\textcolor[rgb]{1,0,0}}
\newcommand{\grad}{\mbox{$^{\circ}$}}
\newcommand{\ksic}{$\overline{\textrm{K}}_{\textup{SiC}}$}
\newcommand{\mau}{$\overline{\textrm{M}}_{\textup{Au}}$}
\newcommand{\kau}{$\overline{\textrm{K}}_{\textup{Au}}$}
\newcommand{\mgr}{$\overline{\textrm{M}}_{\textup{Gr}}$}
\newcommand{\kgr}{$\overline{\textrm{K}}_{\textup{Gr}}$}


\title{Supplementary information for: \\
Semiconductor to metal transition in two-dimensional gold and its van der Waals
heterostack with graphene}

\author{S. Forti}
\email{stiven.forti@iit.it}
\affiliation{Center for Nanotechnology Innovation @ NEST, Istituto Italiano di Tecnologia, Piazza San Silvestro 12, 56127 Pisa, Italy}
\affiliation{Max-Planck-Institut f\"{u}r
Festk\"{o}rperforschung, Heisenbergstr. 1, D-70569 Stuttgart}
\author{S. Link}
\affiliation{Max-Planck-Institut f\"{u}r
Festk\"{o}rperforschung, Heisenbergstr. 1, D-70569 Stuttgart}
\author{A. St\"ohr}
\affiliation{Max-Planck-Institut f\"{u}r
Festk\"{o}rperforschung, Heisenbergstr. 1, D-70569 Stuttgart}
\author{Y. R. Niu}
\affiliation{MAXIV laboratory, Lund University, P.O. Box 118, Lund, S-22100, Sweden}
\author{A. A. Zakharov}
\affiliation{MAXIV laboratory, Lund University, P.O. Box 118, Lund, S-22100, Sweden}
\author{C. Coletti}
\affiliation{Center for Nanotechnology Innovation @ NEST, Istituto Italiano di Tecnologia, Piazza San Silvestro 12, 56127 Pisa, Italy}
\affiliation{Graphene Labs, Istituto Italiano di Tecnologia, via Morego 30, 16163 Genova, Italy}
\author{U. Starke}
\affiliation{Max-Planck-Institut f\"{u}r Festk\"{o}rperforschung,
Heisenbergstr. 1, D-70569 Stuttgart}

\maketitle

\date{\today}


\setcounter{figure}{0}
\renewcommand{\thefigure}{S\arabic{figure}}

\setcounter{section}{0}

\section{Statistical analysis of the 2D ML Au semiconducting character}\label{statanal}
\begin{figure}
\centering
\includegraphics[width=0.75\textwidth]{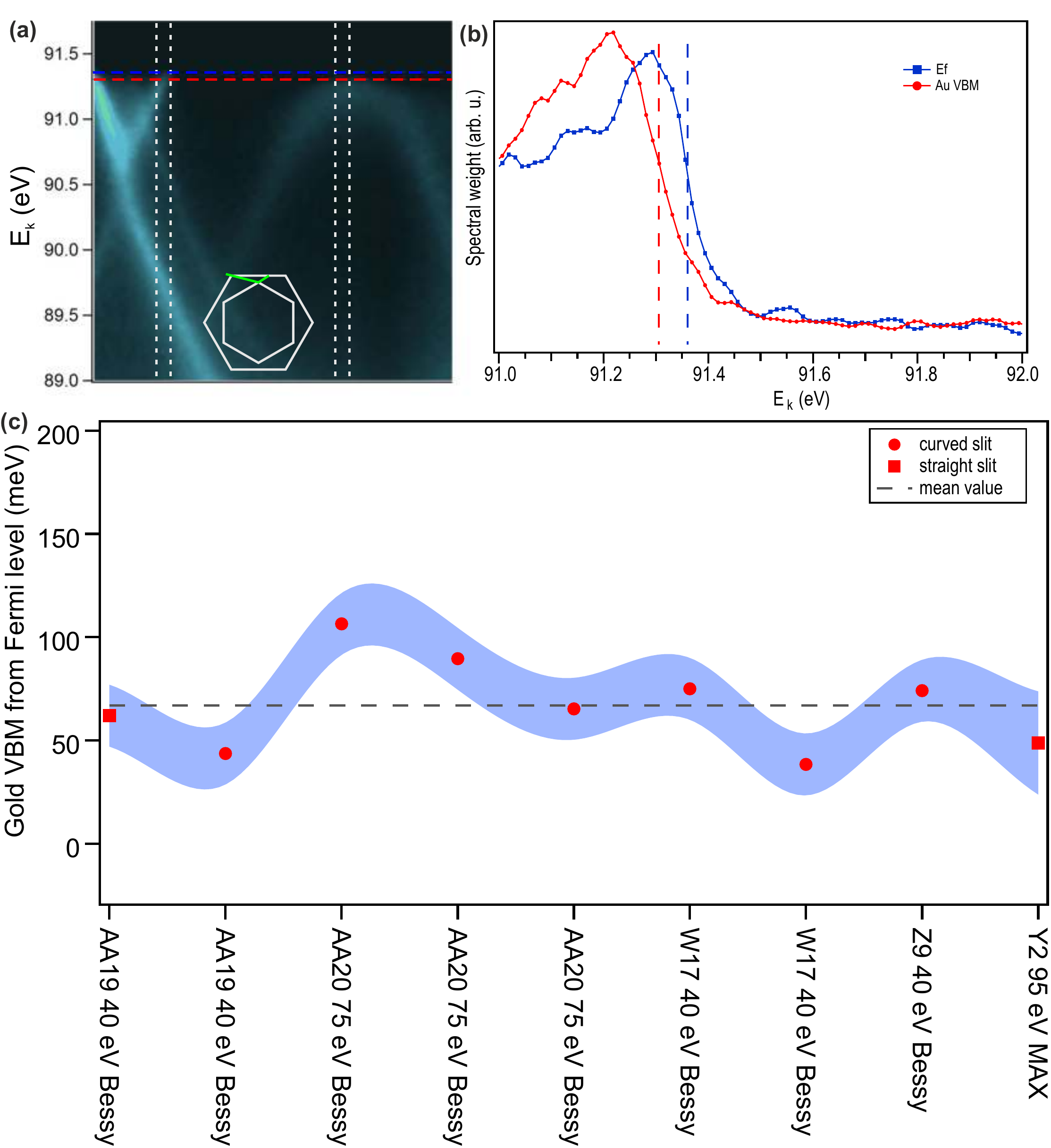}
\caption{(a) ARPES cut along the green line in the inset, showing both the {\kgr} and {\kau} points. The horizontal lines account for the energy level as extracted by the sigmoidal fit of the profiles in panel (b). (b) Profiles taken along the vertical grey-dashed lines in panel (a). (c) distribution of the energy difference between the gold VBM and the Fermi level for different samples.}
\label{FigS1}
\end{figure}

In order to present a robust and statistically solid set of measurements for proving the actual semiconductor character of the single-layer 2D gold, we carried out a systematic analysis of all our ARPES measurements, done on several gold-intercalated graphene samples, at different photon energies and at different synchrotron facilities. Such an extensive analysis allows us to present strong and consistent evidence of the semiconducting behavior of the 2D gold layer. The result of such analysis is reported in Fig. \ref{FigS1}. In panel (a), we show an exemplary ARPES measurement cutting through both the K points of graphene and 2D gold (cf. green line in the inset). In panel (b), we report the energy distribution curves (EDCs) through the gray dashed lines at graphene’s pi bands and at the {\kau} point. The two profiles, i.e. the Fermi edge and the position of the gold valence band maximum (VBM), are fitted with a sigmoidal curve and the fit result is plotted as two vertical dashed lines on the graph. The same procedure was applied for every sample measured and the overall result is reported in the lower part of the figure, where the shaded region represents the error associated with the extracted value.
The $x$ axis of panel (c) shows the names of the samples measured, the photon energy and the facility at which they have been measured. The mean value extracted from this statistics is 67 meV, i.e. the energy distance from the 2D-Au VBM and the Fermi level.

\section{Two-dimensionality of Au states}\label{Gold2D}
\begin{figure}
\centering
\includegraphics[width=0.95\textwidth]{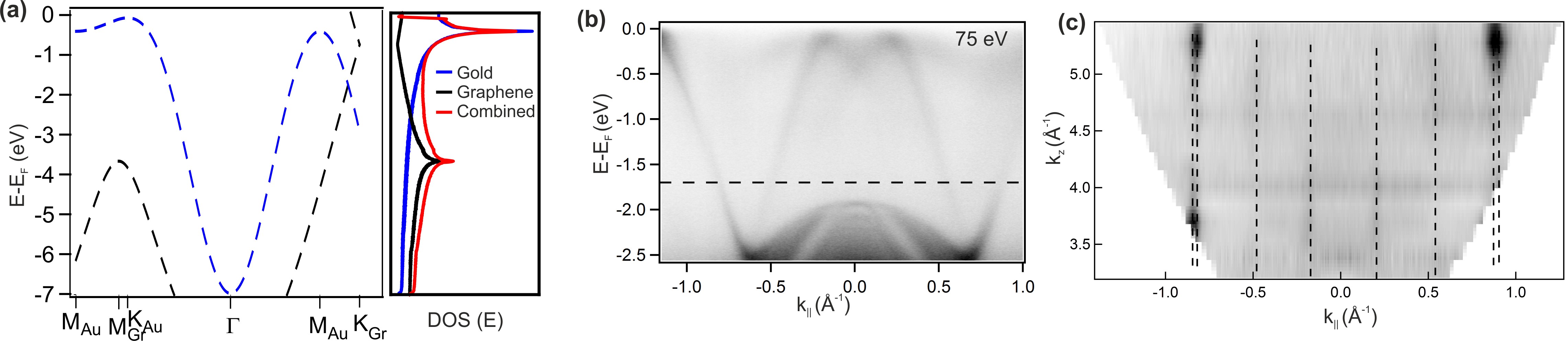}
\caption{(a) NNN-TB bands of gold (blue-dashed) and graphene (black-dashed). In the right panel, the DOS of gold (blue), graphene (black) and the sum of the two (red) is shown in the same energy range. 
(b) ARPES spectrum acquired at 75 eV along the gold {\kp\gp\kpp} direction. (c) $k_z$ dispersion obtained by extracting the spectral weight of spectra similar to the one in panel (a) at 1.7 eV (see dashed line), recorded at photon energies from 30 to 100 eV with 5 eV of energy step.}
\label{FigS2}
\end{figure}

For completeness of what is reported in the main text, in Fig.~\ref{FigS2}(a) we show the TB bands calculated in the NNN approximation on a 2D triangular lattice. In panel (a), on the right side we show the density of states (DOS) derived from the TB model. The 2D-Au exhibits a Van Hove singularity at its {\mpp} point, which is located just about 400 meV from the Fermi level. This means that this system has an instability point which is at an energy reachable by electrical gating. 
To prove the two-dimensionality of the gold bands, we show high-resolution ARPES data, collected with different photon energies to measure the dispersion of the $k$-vector along the $z$ direction in reciprocal space.
In panel (b) we show an exemplary of ARPES spectrum acquired along the Au {\kp\gp\kpp} direction and centered in {\gp} with photon energy 75 eV. The spectral weight visible in the vicinity of {\gp} is due to gold replica bands as it will be further explained in sec. \ref{replicas}. The spectral weight at binding energy 1.7 eV, as traced by the dashed line in the panel, was extracted for every energy from 30 to 100 eV with energy step 5 eV and plotted as shown in panel (c), following the relation that binds the $k_z$ vector to the kinetik energy and the emission angle: $k_{z} = \sqrt{\frac{2m}{\hbar}}\sqrt{E_{kin}\cos^2(\theta)+V_0}$, where $V_0$ is the \textit{inner potential} and $\theta$ the photoemission angle. We chose $V_0=14.5$ eV, according to Refs.~\cite{ZhouAnnPhys2006,OhtaPRL2007} and considering the difference in the valence band minimum between gold and graphene as extracted from the TB calculations. The absence of dispersion in the direction perpendicular to the surface of the $k$-vector indicates the actual two-dimensionality of the interfacial gold layer. The spectral weight visible towards the Fermi level in {\gp} is ascribed to gold replica bands, due to electron diffraction at the surface. The scattering vector corresponds to the reciprocal lattice vector of graphene, hence rotated by 30 degrees with respect to the gold BZ alignment (see also sec. \ref{replicas}).

\section{The electronic properties of graphene over a single layer of gold}\label{manybody}

\begin{figure}
\centering
\includegraphics[width=0.95\textwidth]{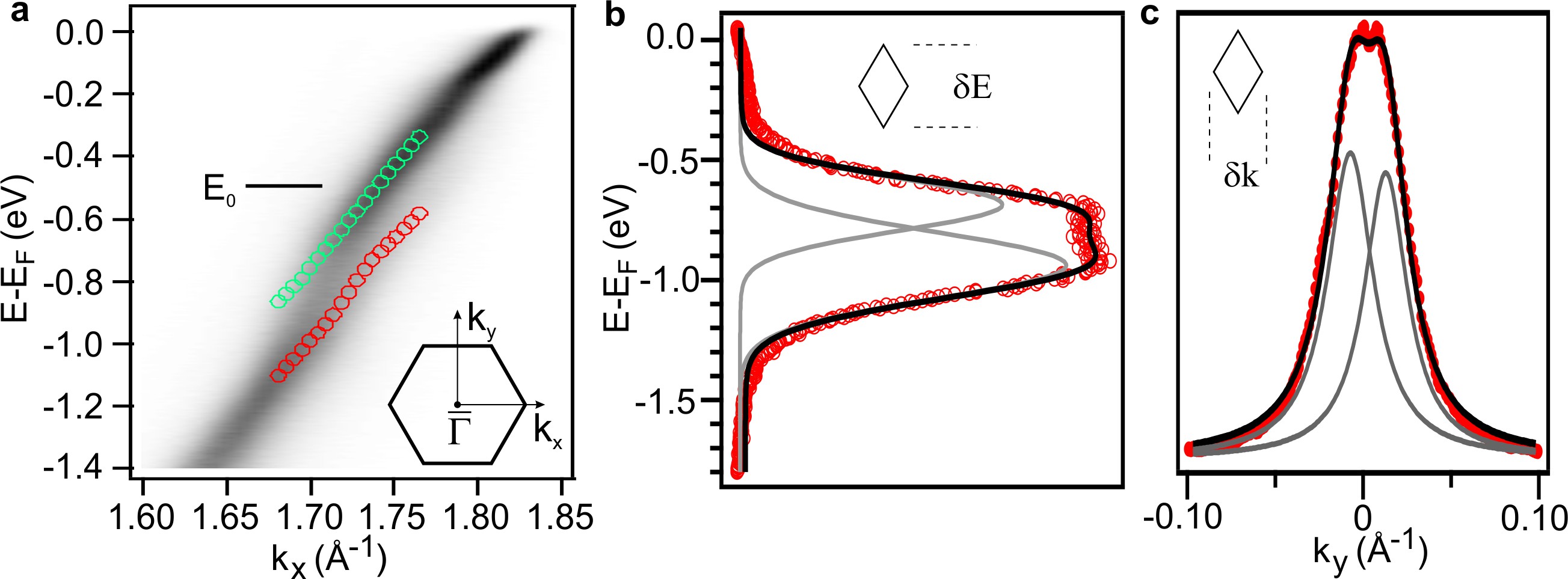}
\caption{{\bf Observed plasmarons in graphene.} (a) ARPES spectrum at 40 eV of the Dirac cone along the $k_x$ direction of a Au-intercalated epitaxial graphene, $n$-phase. The empty circles are fits of the peak position of the energy distribution curves. (b) and (c) line profiles of the plasmaron diamond, from which to extract the value of dielectric constant. See also Fig.~\ref{FigS4}.}
\label{FigS3}
\end{figure}

\begin{figure}[h!]
\centering
\includegraphics[width=0.95\textwidth]{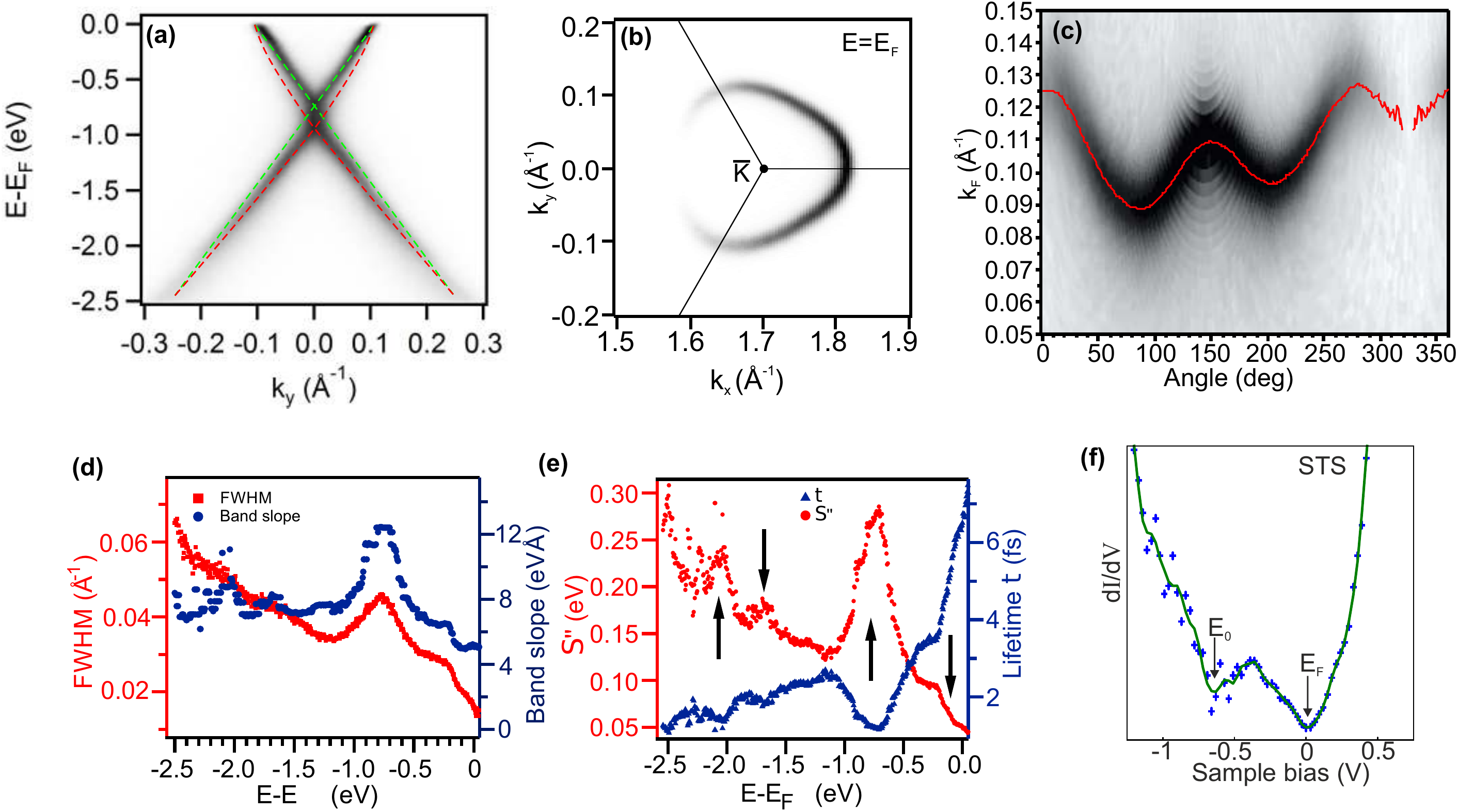}
\caption{(a) ARPES spectrum of the graphene's Dirac cone measured perpendicular to the {\gp\kgr} direction at 40 eV. Photohole and plasmaron dispersion are superimposed to guide the eye. (b) Fermi surface of graphene. (c) Angular dependence of the Fermi momentum. (d) FWHM and band slope of the Dirac cone extracted along the {\gp\kgr} direction at 40 eV. (e) Imaginary part of the self energy and quasiparticle lifetime extracted from the FWHM (see text). (f) Scanning tunneling spectroscopy of Gr/SC Au/SiC(0001) acquired at 78 K. $E_0$ is the Dirac energy, as defined in the main text.}
\label{FigS4}
\end{figure}

The effects of gold intercalation over the decoupled monolayer graphene on SiC(0001) have been already discussed in the past literature\cite{GierzPRB2010, WalterPRB2011}.
Nevertheless, in this section we point out what are the effects of a very careful preparation on the properties of the intercalated graphene.\\
The very careful preparation of the samples via the intercalation of precisely one Au monolayer with respect to the SiC(0001) atomic areal density, translates in a very clean ARPES signal and allows for the extraction of band parameters from the data. Fig.~\ref{FigS3}(a) shows the ARPES spectrum of the Dirac cone acquired along {\gp\kgr} at 40 eV. The measured bands are accompained by the fitting of the positions of the maxima of the energy dispersion curves, shown as green and red circles superimposed to the raw data. Two distinct dispersing states are fitted, as it is apparent from the figure. This effect has been reported for the first time in graphene by Bostwick and coworkers~\cite{BostwickScience2010}. What is visible below the Dirac point is the energy dispersion of the \textit{plasmaron} quasiparticle\cite{LundqvistPKM1969}, i.e. a photohole that is coupled to a plasmon with similar group velocity. According to Ref.~\citep{WalterPRB2011}, by measuring the energy and momentum separation of the hole and plasmaron bands, one can estimate the effective dielectric constant of the investigated sample. We determined the energy and momentum spread of the diamond formed by the crossing of the hole and the plasmaron dispersions (cf. Fig.~\ref{FigS4}(a)). This is done in Fig.~\ref{FigS3}(b) and (c), where the energy (b) and momentum (c) distribution curves of the Dirac cone are fitted with two Voigt functions, colored in gray. The two band traces are about 253 meV (b) and 0.02 {\AA}$^{-1}$ (c) apart from each other and the  Dirac energy of the system is defined as E$_0$ and located (685$\pm$5) meV below the Fermi level (cf. Fig.~\ref{FigS4}). 
To give an estimation of the effective coupling constant and effective dielectric constant, we compare our results with the work of Ref.~\onlinecite{WalterPRB2011}. However, we point out that the normalization of the diamond width in momentum space with respect to the Fermi momentum $k_F$ is somewhat inconsistent since for such high doping level, the Dirac cone is strongly warped (see Fig. \ref{FigS4}(b) and (c)). We therefore rely on the separation of the cones in energy. In this way, the $\alpha_{ee}$, or graphene fine constant is found to be $\alpha_{ee}=0.30\pm0.02$, which is translated into an effective dielectric constant $\epsilon_{eff}=7\pm1$, meaning a substrate dielectric constant of $\epsilon_s=2*\epsilon_{eff}-1=13\pm2$.

Such a small effective dielectric constant enhances the \textit{e-e} coupling and makes collective phenomena such as plasmarons observable.

In their paper\cite{WalterPRB2011}, Walter and coworkers reported a dielectric constant value about five times higher. This is simply because they started from a $p$-type Au intercalated graphene, which is induced by twice the amount of gold at the interface (cf. Fig.~3(c)), and counter $n$-doped it with K atoms in order to set the Fermi level of the system well above the Dirac point and thereby observe the plasmaron band.

For completeness, we also extract the band parameters from the measured ARPES. To do this, we analyze the ARPES data of the graphene Dirac cone measured at 40 eV, as displayed in Fig.~\ref{FigS4}. In panel (a) we show the ARPES spectrum of the $n$-doped Dirac cone as in Fig.~1 of the main text, with the addition of dashed lines to distinguish between the hole (green) and plasmaron (red) dispersions. Panel (b) displays the Fermi surface, whereas panel (c) shows the amplitude of the Fermi vector as a function of the angle as centered in {\kp}, highlighting the threefold symmetry of the Fermi surface and the dark corridor\cite{ShirleyPRB1995}, where the intensity vanishes. This supports the argumentation promulgated in the main text, that considers unreliable the extraction of the dielectric constant value by measuring the distance in $k$-space between the hole and plasmaron dispersion, normalized for the Fermi vector. Such a normalization procedure is clearly direction-dependent, instead of the energy difference between the two dispersions, which is not, and it is therefore more reliable. 
We extracted the band parameters by fitting the momentum dispersion curves (MDCs) with Lorentzian lineshapes. Fig.~\ref{FigS4}(d) shows the full width at half maximum (FWHM) of the bands and their slope $\partial E/\partial k$, in the energy range displayed in panel (a). Panel (e) shows the imaginary component of the self energy and the photohole lifetime, which depend on the FWHM through the relation: $\Sigma^{''}=(\delta k/2)(\hbar v_F)$ and $\tau=1/(v_F \delta k)$, where $\delta k$ is the FWHM and $v_F$ the Fermi velocity. The self energy is a particularly sensitive quantity for measuring the interactions between quasiparticles. Indeed, a strong peak is visible in proximity of the Dirac point, corresponding to the photohole relaxation through the emission of a plasmon. The other two less pronounced peaks are ascribed to the interaction between graphene and gold bands, although in the dispersion no actual gap is observed. The first low-energy kink in the spectrum is due instead to quasiparticle relaxation via phonon emission. Fig.~\ref{FigS4}(f) displays the scanning tunneling spectroscopy (STS) spectrum of the semiconducting Au phase intercalated in between graphene and SiC(0001), acquired at 78 K. The Dirac point is observed at around -625 meV, in line with the ARPES measurements. 

\section{Interfacial atomic ordering derived from periodicity of ARPES replica bands}\label{replicas}
\begin{figure}[t!]
\centering
\includegraphics[width=0.95\textwidth]{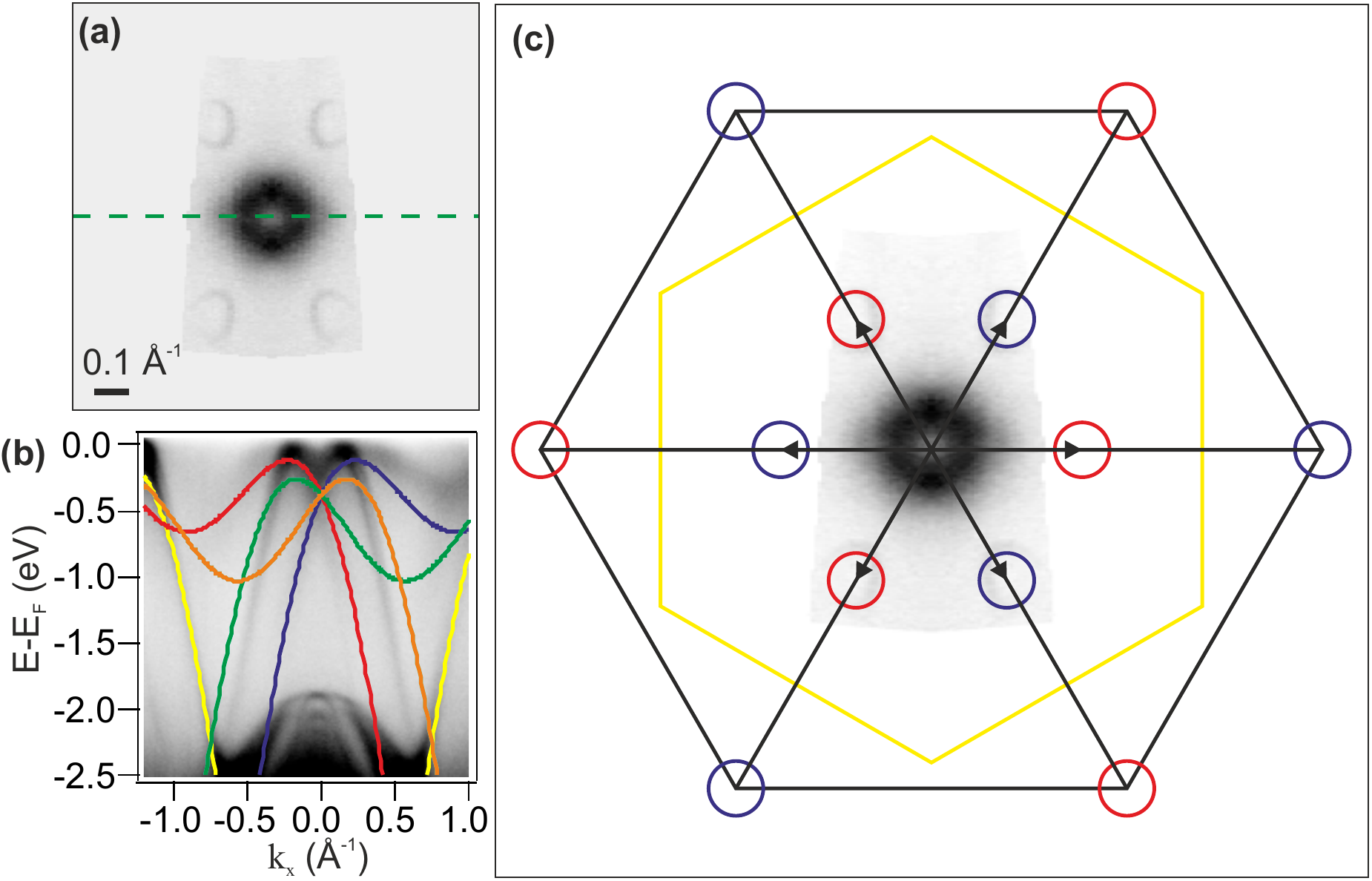}
\caption{(a) ARPES constant energy surface close to the Fermi level measured about {\gp} at 40 eV. (b) Model for the SC gold replica bands, imposing the graphene reciprocal lattice vector as generator. (c) Model for the graphene replica bands using a SC Au reciprocal lattice vector as generator.}
\label{FigS5}
\end{figure}

In this section we show the final state effects caused by the diffraction of photoemitted electrons as they exit the material.\\
In Fig.~\ref{FigS5}(a) a portion of the $k$-space is shown, recorded with ARPES at 40 eV centered in {\gp}. At this binding energy, no intensity should be observable. However, some spectral weight clearly appears. Four Dirac cone profiles are distinguishable and correspond to graphene replica bands. Those replicas are induced by a reciprocal lattice vector with module and orientation corresponding to the SiC(0001) surface vector. They are therefore generated by the gold atoms arranged on the SiC (1$\times$1), as shown in panel (c).
The intensity around {\gp} is due to gold replica bands. In this case the generator vector is the reciprocal lattice vector of graphene. In panel (b) we show an ARPES spectrum recorded at 75 eV along the gold {\kp\gp\kpp} direction with superimposed the theoretical gold SC bands derived from TB (in yellow) and displaced by a graphene reciprocal lattice vector (other colors). The matching is quite remarkable, confirming the proposed model and also justifying the residual intensity visible in the $k_z$ vs. $E$ plot in Fig.~\ref{FigS2}(c).

\section{Details about HRXPS data fitting}
\label{SecXPS}
In this section we report the details about the fitting for the HRXPS data, as shown in Fig.~2 of the main text. \\ 
The Au~4f spectrum of the 2D SC phase can be well fitted with a single Voigt doublet lineshape (cf. Fig.~2(a)). The Si~2p spectrum instead, needs a double Voigt doublet function (cf. Fig. 2(b)). For the Au~4f, spin splitting was set to 3.68~eV and the branching ratio was 0.75. Standard constraints were set for the Si~2p fit, such as spin splitting of 0.63~eV and branching ratio of 0.5 for the single doublets. Within the respective doublets, both components were set to the same Lorentzian and Gaussian widths. The resulting fitting parameters can be found in Tab.~\ref{Tab1}.
\begin{table}[htb] \centering
	\begin{tabular}[b]{l c c c c}
		& E$_{\textup{B}}$ (eV) & $\omega_\textup{L}$ (eV) & $\omega_\textup{G}$ (eV) & ratio (\%)\\
		Au~4f$_{\sfrac{7}{2}}$ & 84.35 & 0.15 & 0.20 & \\
		Si~2p$_{\sfrac{3}{2}}$, bulk SiC & 100.78 & 0.12 & 0.32 & 68\\
		Si~2p$_{\sfrac{3}{2}}$, Au-Si & 100.31 & 0.12 & 0.13 & 32\\
	\end{tabular}\caption{Fit parameters of the Si~2p$_{\textup{\sfrac{3}{2}}}$ and the Au~4f$_{\sfrac{7}{2}}$ components of the n-phase Au intercalated ZLG. The binding energies of the respective doublet partners are given by the spin splitting mentioned in the text, which was set as a constrain for the fit.} \label{Tab1}
\end{table}
The small Lorentzian and Gaussian widths for the Au~4f levels is indicative for the presence of only one chemical species of Au. The 4f$_{\sfrac{7}{2}}$ component's binding energy of 84.35~eV is shifted by about 350~meV towards higher binding energy compared to metallic Au. This means firstly, that one can exclude the presence of Au clusters on the sample surface, like observed earlier~\cite{GierzPRB2010}. Also, all Au atoms within the interface have the same chemical environment. Judging from the two very distinct chemical components in the Si~2p spectrum, it is reasonable to assume that these Au atoms are chemically bound to the topmost Si of the SiC(0001) surface. All components of the Si~2p spectrum show a Lorentzian broadening of 0.12~eV, which can be viewed standard for this core level. The Gaussian width is 0.32~eV in the 100.78~eV binding energy doublet and 0.13~eV in the 100.31~eV doublet (both binding energies denote to the 2p$_{ \sfrac{3}{2}}$ components). A Gaussian broadening of 0.32~eV is generally observed for bulk SiC even if the experimental resolution is much better. However, the Gaussian width of Si~2p spectra can be much smaller in other materials, as it is the case for instance in the (7$\times$7)-reconstruction of the Si(111) surface~\cite{PaggelPRB1994}. This peculiar broadness within the SiC system is not understood up until now, yet it can be seen as characteristic for bulk SiC. It is therefore reasonable to assign the narrower component to Si, which is bound to Au on the surface of the SiC. Au adsorbed on the (0001) surface of 4H-SiC have been investigated by Stoltz and coworkers\cite{StoltzJPCM2007}. They report a Si~2p component distribution, which is similar to our case, but with slightly different binding energies. There, they also assign the higher binding energy component to Si in the bulk of the SiC and the lower binding energy component to Si, which is bound to Au on the surface. The differences in binding energy can be caused by the different SiC polytype used here (6H-SiC) and by them (4H-SiC). Also the additional graphene layer on top in our case has influence on the band alignment and accompanied surface band bending in the SiC~\cite{RisteinPRL2012,Huefner}, which alters the observed binding energies.\\
It should be noted, that one can compare the intensity ratio of the signal produced by the two chemical Si species to other intercalation systems based on graphene on SiC(0001). The intensity ratio from Si in the bulk SiC and Si bound to Au is 68:32. In the case of H intercalation, the intensity ratio of Si in the bulk SiC to Si bound to H is 63:37 and 67:33 for photon energies 140~eV and 330~eV, respectively~\cite{RiedlDissertation}. In this case, every Si atom on the surface of the SiC is bound to one H. Keeping in mind that the photon energy in the measurements presented here is 210~eV, it is reasonable to assume that also in the Au case here all Si atoms on the surface are bound to Au.\\
\begin{figure}
\centering
\includegraphics[width=0.95\textwidth]{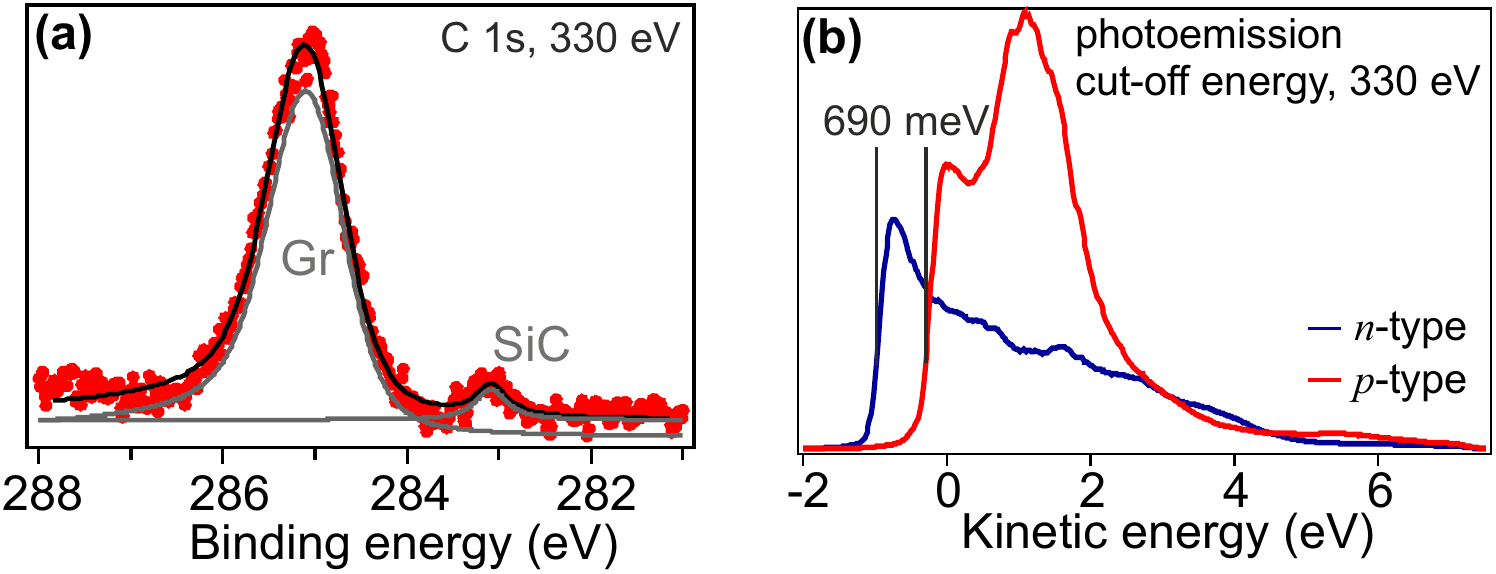}
\caption{(a) $\mu$XPS spectrum of C 1s recorded on Gr/SC 2D Au/SiC(0001) at 330 eV. (b) Photoemission cut-off spectra recorded on Gr/Au/SiC(0001) regions where the graphene was $n$ and $p$ doped.}
\label{FigS6}
\end{figure}
The decoupling of the graphene is well illustrated by the C 1s core level peak in Fig.~\ref{FigS6}(a). The SiC peak binding energy is measured at about 283 eV, as observed in other intercalated systems\cite{RiedlPRL2009,FortiPRB2011,Forti2DMat2016}. The graphene peak is here fitted with a single Doniach-{\^S}unji{\'c} (DS) centered at 285 eV with a gaussian width of 0.45 eV, which takes into account the different chemical environments present inside the supercell\cite{Forti2DMat2016,PreobrajenskiPRB2008}. The binding energy of the sp$^2$ carbon peak is increased by about 600 meV from the nominal peak position of 284.4 eV. Such a shift is mostly due to the $n$-type doping of about $n\sim0.035$ electrons per graphene unit cell\cite{Schroeder2DMater2016}. In  turn, we observe this doping being accompanied by a decrease of work function. Being a 2D material, whenever the bands are rigidly moved up or down with respect to the Fermi level, a corresponding change in the work function must occur. This is actually observed and measured through the cut-off energy of the photoemitted electrons, displayed in Fig.~\ref{FigS6}(b). We look at the low-kinetic energy part of the photoemission spectrum until we reach the edge where no electron is extracted from the material. In this way we can easily measure the difference in work function between regions which are $n$ and $p$ doped (corresponding to semiconducting and metallic gold phases, respectively). We measure an energy difference of (690$\pm$5) meV. Considering that the $p$-phase is almost neutral (cf. Fig.~3 in the main text), such a measure shows very well how tightly related the charge transfer and the variation of work function are in 2D materials. \\

\section{STM measurements}\label{STMsec}
\begin{figure}
\centering
\includegraphics[width=0.95\textwidth]{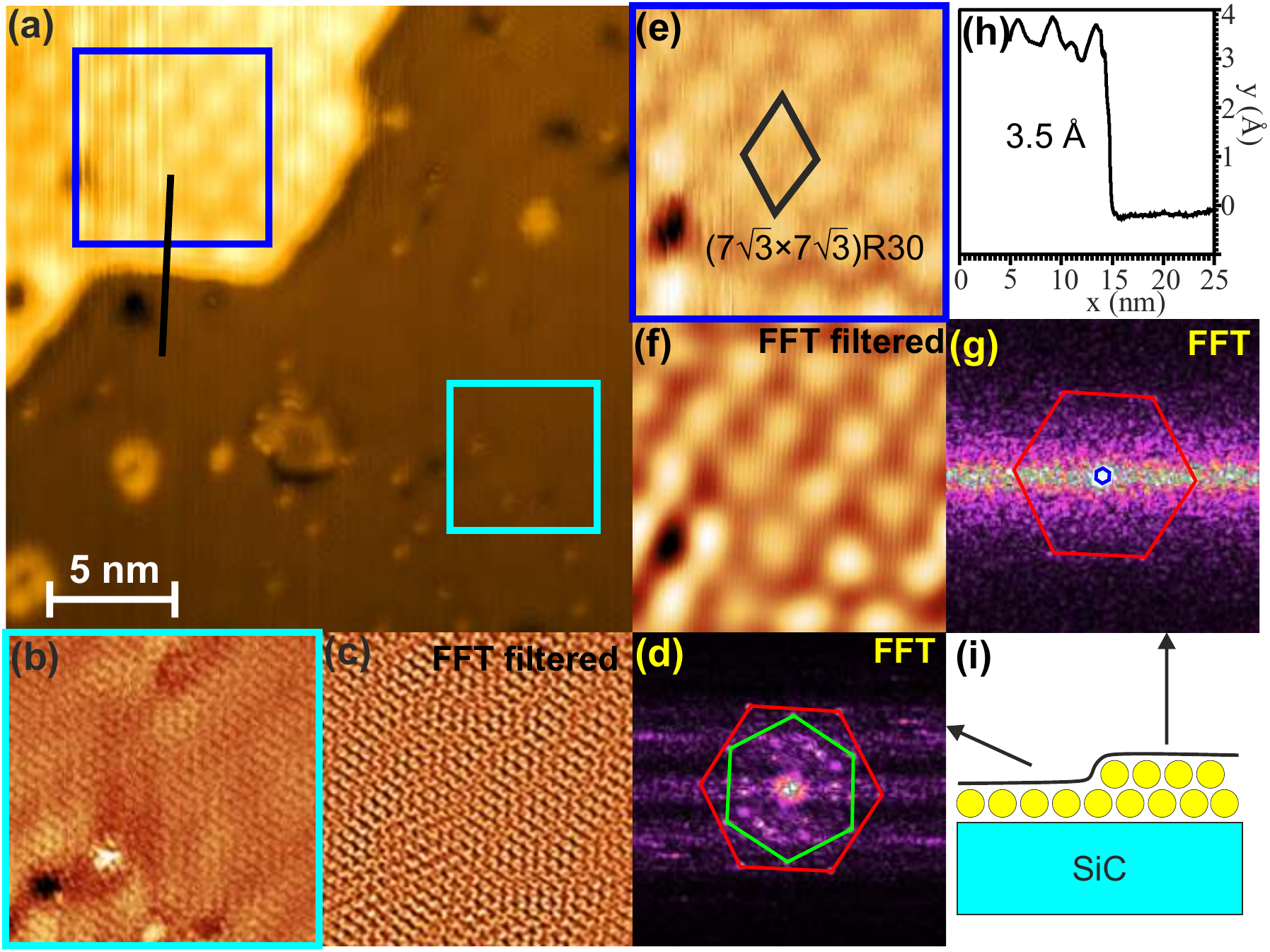}
\caption{(a) STM topography of a SC/M transition region. Au-intercalated graphene recorded with 400 pA of constant tunnel current and -250 mV of tip voltage. Panels (b,c,d) and (e,f,g) show a zoomed-in portion of the scanned area corresponding to the frames on panel (a), the FFT filtered image of the same region and the 2D-FFT of the image of the SC and M regions, respectively. Panel (h) shows the black line profile across the transition region between the SC and the M regions in panel (a). (i) simple sketch of the system.}
\label{FigS7}
\end{figure}

\begin{figure}[t!]
\centering
\includegraphics[width=0.95\textwidth]{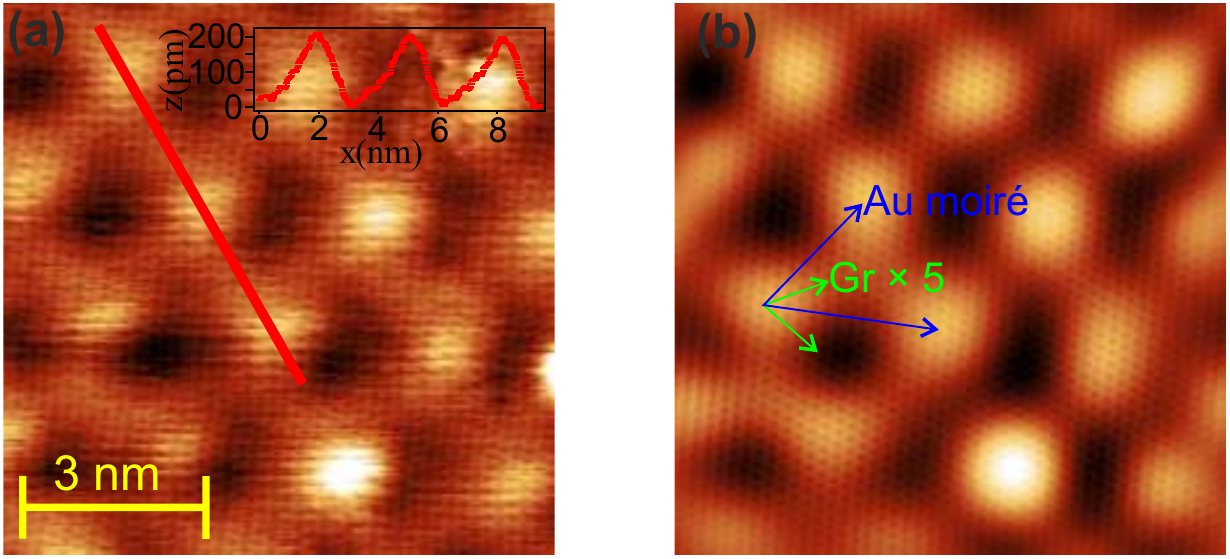}
\caption{{\bf M-phase superstructure measure with STM.} (a) Raw topographical data. In the top-right inset, the line profile over the red line is shown. (b) 2D-FFT filtered image with indicated the lattice vectors of the superperiodicity and of graphene, the length of which has been increased by a factor 5.}
\label{FigS8}
\end{figure}

In this section, we show scanning tunneling microscopy (STM) measurements of the gold-intercalated graphene in its SC and M phase.
Fig.~\ref{FigS7}(a) shows a sample's region where the transition between the SC and M gold region is visible. The SC (M) region is shown zoomed-in in panel b (e), whereas the FFT-filtered and 2D-FFT of the zoomed-in region are shown in panel c and d (f and g), respectively. The superstructure on the M region is so large and strong that it can be seen even on the large-scale image. The corrugation of the M region pattern is indeed of the order of 200 pm, as visible also in the line profile of panel (h) (cf. also Fig.~\ref{FigS8}), while on the SC phase is about 25 pm. 
As visible from panel (e) and Fig.~\ref{FigS8}, the superstructure stemming from the M region has a periodicity of  (7$\sqrt{3}\times$7$\sqrt{3}$)R30 graphene unit cells over (10$\times$10) gold, as also confirmed by $\mu$LEED measurements (see discussion in the main text).
The height difference of about 3.5 {\AA} observed between the two regions supports a model where only two gold layers are intercalated in the M phase.

\begin{figure}
\centering
\includegraphics[width=0.95\textwidth]{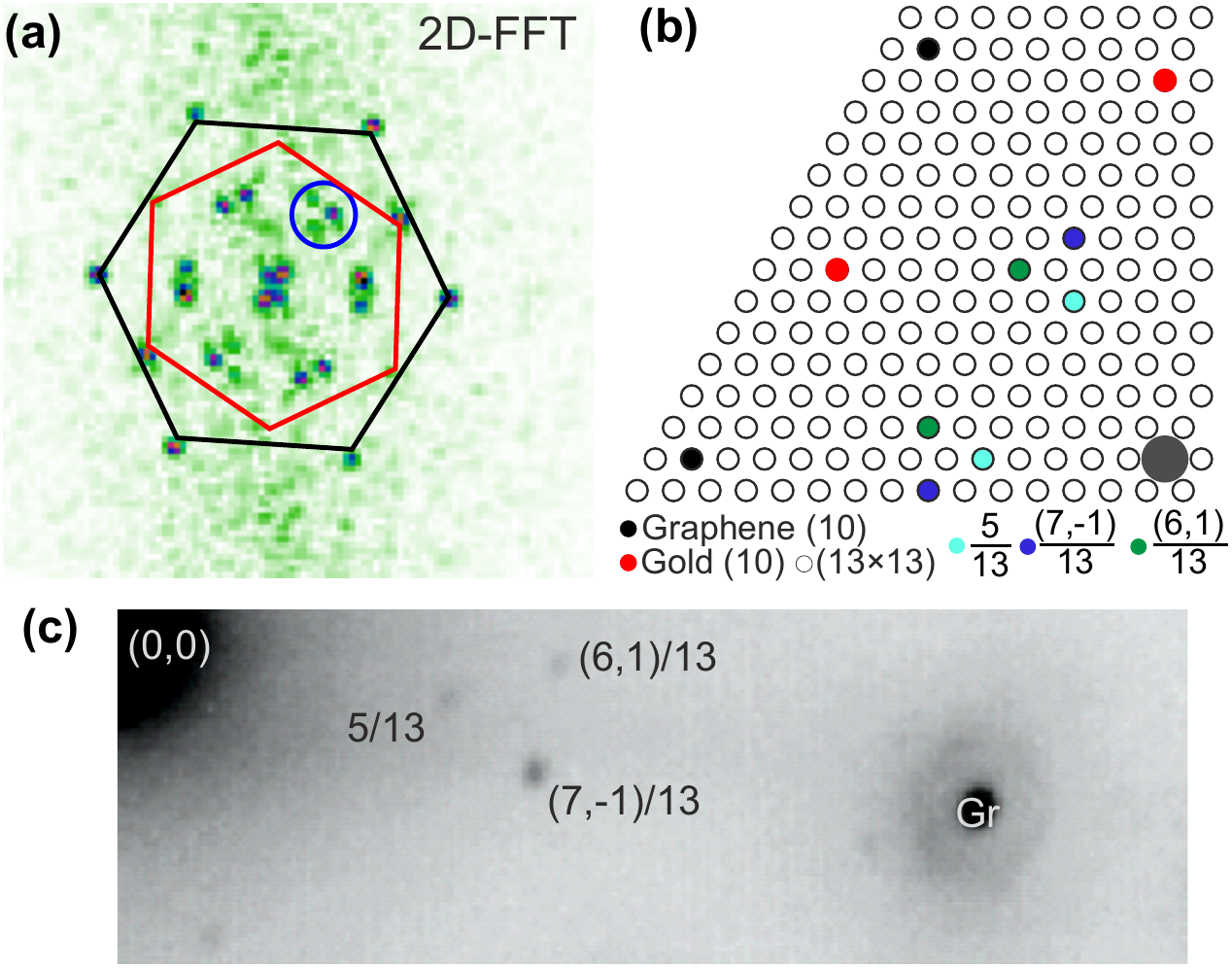}
\caption{(a) 2D-FFT of SC Au phase, retrieved from Fig.~2(d). Graphene and gold reciprocal lattice spots are indicated by a black and red hexagon, respectively. The spots stemming from the (13$\times$13) periodicity are circled in blue. (b) Theoretical reciprocal lattice grid of the (13$\times$13), representing the kinematic LEED pattern in the first quadrant. Highlighted are the spots visible in panel (a). (c) portion of Fig. 2(c) from the main text with enhanced contrast, to highlight the high order diffraction spots on the (13$\times$13) grid.}
\label{FigS9}
\end{figure}
The 2D fast Fourier transformed (FFT) image FFT pattern in Fig.~\ref{FigS9}(a) clearly shows the graphene and gold reciprocal lattice spots, indicated with a black and red hexagon, respectively. Other spots are also visible and they belong to the (13$\times$13) grid, as explained in panel (b), where the top-left quadrant of the theoretical LEED pattern up to the graphene's first diffraction order is shown. The blue circle in panel (a) encloses three spots, two of which are visible along every direction. The third and innermost spot is less visible in the FFT. As illustrated in panel (b), those spots are the (7,-1)/13, (6,1)/13 and 5/13 reflexes of the (13$\times$13) pattern. The same spots are measured as well by $\mu$LEED, as we prove by showing in panel (c) a portion of Fig. 2(c) with enhanced contrast. 
We briefly point out that the (13$\times$13) is a rather natural periodicity for the graphene on SiC system and it is often observed, also for other intercalated systems\cite{Forti2DMat2016}. In Ref.\cite{PremlalAPL2009} , for example, they have observed several different superperiodicities induced by the intercalated gold. Even the (13$\times$13), but in that case, the gold was aligned with the graphene and it had a different lattice constant. A configuration very similar to what has been observed for the copper-intercalated graphene\cite{Forti2DMat2016, KazumaAPL2014}.

\newpage

\section*{References}
\bibliographystyle{./SFbibstyle_NPG}
\bibliography{./Biblio}